# Structural, magnetic, and critical behavior of CrTe$_{1-x}$Sb$_x$ Alloys

M. Kh. Hamad[1*], Yazan Maswadeh[2], E. Martinez-Teran[3], A. A. El-Gendy[3*,] and Kh. A. Ziq[1*]

[1]Department of Physics, King Fahd University of Petroleum & Minerals, Dhahran 31261, Saudi Arabia

[2]Department of Physics and Science of Advanced Materials Program, Central Michigan University, Mt. Pleasant, Michigan 48859, USA.

[3]Department of Physics, University of Texas at El Paso, TX79968, USA

**Abstract**

We investigate the structural and critical properties of *CrTe$_{1-x}$Sb$_x$* with $0.0 \leq x \leq 0.2$. The XRD patterns revealed that Sb-substitution resulted in a pure *NiAs*-hexagonal structure with P6$_3$/mmc (194) space-group. Lattice refinement of the structure revealed little changes in the *a*-lattice parameter, along with a more pronounced reduction in the *c-axis*. The critical behavior in *CrTe$_{1-x}$Sb$_x$* has been investigated using the magnetization isotherms near the ferromagnetic transition. The obtained critical exponents ($\beta, \gamma,$ and $\delta$) revealed that all samples (with $0.0 \leq x \leq 0.2$) closely follow a mean field-like behavior with ferromagnetic Curie temperature (Tc) near room temperature. The results from Widom scaling relation indicating self-consistency of the acquired values. Moreover, the magnetization isotherms near the Curie temperature follow a universal scaling behavior, giving further support for the obtained critical exponents.

[*]**Corresponding authors**: mhamad@kfupm.edu.sa;  aelgendy@utep.edu;  kaziq@kfupm.edu.sa

## 1- Introduction

Cr-based pnictogens and chalcogens continue to attract the attention of scientists due to their versatile physical properties [1-5]. These properties range from superconducting to semiconducting and half-metallic [6-7], ferromagnetic van der Waals to antiferromagnetic (AFM) behavior [8-14]. Some of the emerging possible applications are in magnetic cooling and spintronic devices [14]. The binary compounds of Cr-pnictogens  (N, As, Sb) and the Cr-chalcogens (S, Se, Te) crystalize in the hexagonal NiAs crystal structure [1-5], however the chemical bonding may lead to diverse magnetic and physical properties. Ionic bonding dominates the chalcogenides, while the bonding in pnictide is non-ionic. For instance, $Cr_xTe$ is a ferromagnetic (FM) material with a Curie temperature covering wide range (320 – 360 K) depending on the value of x [6-15]. It is well established that $Cr_xTe$ is non-stoichiometric and does not occur in pure-hexagonal phase [16-18]. While substitution at the Cr sites helps to stabilize the hexagonal NiAs structure. Deficiency and defects may also stabilize other structures like trigonal $Cr_2Te_3$ and monoclinic $Cr_3Te_4$ structure. Chromium antimony CrSb is antiferromagnetic with $T_N$ ~ 700K, it crystalizes in a NiAs-type structure [19-23] with $\mu_{eff}$ = 3.89 $\mu_B$ [24, 25]. In CrSb; the Cr spins are aligned ferromagnetically within the basal plane; while in the adjacent planes the spins are aligned antiferromagnetically [23]. Substitution in the solid solutions of $CrTe_{1-x}Sb_x$ affects the spins configurations and the magnetic phase in this compound [22, 26]. Increasing the Sb-concentrations; in $CrTe_{1-x}Sb_x$ leads to a gradual reduction in FM ordering temperature and the saturated magnetization along with moderate reduction in unit cell volume. At high concentration of Sb, a more complex magnetic phases may occur such as canted spin structures, noncollinear spin configurations and even spin-glass like behavior have been observed [14, 27-28]. Some of these magnetic changes qualitatively agree with de Gennes model of the double exchange interaction in AFM-FM system [29].

Preliminary critical behavior analysis in $CrTe_{1-x}Sb_x$ has been published in an earlier work [30] where we investigate the effect of Sb substitution over relatively wide range. The modified Arrott plots and Kouvel–Fisher critical exponents' analyses revealed that upon increasing the Sb, the critical exponents deviate significantly from the mean field values and gradually shifts towards 3D Heisenberg models. More recently, we found that repeated grinding and annealing during sample preparation improves the phase purity in $CrTe_{1-x}Sb_x$ especially at low Sb-concentration [14]. For these samples, we investigated the magnetocaloric effect near room temperature $CrTe_{1-x}Sb_x$ and found that the relative cooling power is comparable to pure Gd- the prototype magnetocaloric material.

In this work, we extend our investigation of the magnetic and the structural properties of $CrTe_{1-x}Sb_x$ alloys at low Sb concentrations. The samples used in this study are the same samples we used in Ref. 14. We also investigate the detailed properties of the critical behavior near the FM-PM transition in these newly prepared samples and compare them with our earlier woks [30]. We also include evidences on the changes in the bond length and angle with Sb-substitution that ultimately may affect the magnetic and critical behavior. This may lead to a better understanding of the observed differences with earlier work on the critical behavior [30] in this material.

## 2- Experimental Techniques

Polycrystalline *$CrTe_{1-x}Sb_x$* (*x≤0.2*) were prepared using high temperature solid-state reaction [31-32]. High purity (4-5N) powders of Cr, Te, and Sb are used. The detail of the preparation method is reported in our previous work [30]; however, for the samples in this study we used several repetition of intermediate grinding and annealing to improve the phase purity and quality of the samples. Bruker X-ray diffractometer D2-Phaser with Cu Kα (λ=1.54056 Å) has been used to obtain the powdered XRD patterns over 20-80º [14, 33]. We used Rietveld refinements available in FULLPROF software to analyze the XRD patterns. Quantum design 3-Tesla VersaLab has been used to collect the magnetic data in the temperature range of 50-400K.

## 3- Analysis and Results

Figure 1 shows the room temperature XRD patterns of the studied $CrTe_{1-x}Sb_x$ samples. Rietveld refinement of the data was carried out to determine the lattice parameters, phase purity, and structure [34, 35]. The results and the lattice parameters for all investigated samples are discussed in details elsewhere [14]. Those observations are consistent with the noticed shifts of the positions of the diffractions peaks with the increasing the Sb substitution ration in the samples (Fig. 2.), the color-map at Figure 2 shows a noticeable shift in the (0 0 l) peaks to higher angles indicating a shortening in the *c-lattice* parameter in direct space.

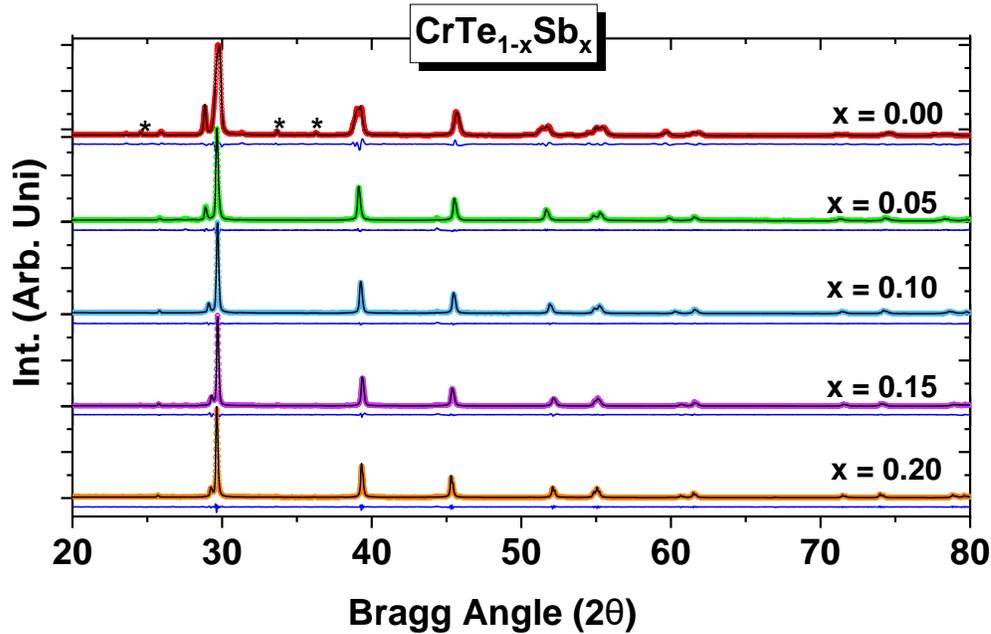

*Figure 1: XRD patterns for $CrTe_{1-x}Sb_x$. The stars indicate the position of the impurity phase.*

A relatively smaller shift to the left in the (*h k 0*) peaks, that indicates a relaxation along the basal plane *a*-lattice. Moreover, the figure (for x=0.0) also reveals the presence of secondary-phase $Cr_2O_3$, which has a trigonal structure with R-3c space group, leading to the nonstoichiometric *$Cr_xTe$* sample in line with earlier results [8]. *Street et al*, concluded that Chromium-telluride *$Cr_xTe$* exists over a range of Te-concentration with the hexagonal structure [8, 36]. The Wyckoff positions for the hexagonal structure are given in Table 1. Sb-substitution causes relaxation of the atomic-bond along (*a-b*) axes, this probably will be reflected residual stress along the *c-axis* (Fig. 3) leading to the observed variations in the lattice parameters and the overall increase in the volume of the unit cell for x=0-0.10 [see 14].

Table 1: The Wyckoff position for $CrTe_{1-x}Sb_x$ hexagonal structure space group $P6_3/mmc$ (194).

| Atom | Wyckoff | Site | x/a | y/b | z/c |
|---|---|---|---|---|---|
| Cr | 2a | -3m. | 0 | 0 | 0 |
| Te/Sb | 2c | -6m2 | 1/3 | 2/3 | 1/4 |

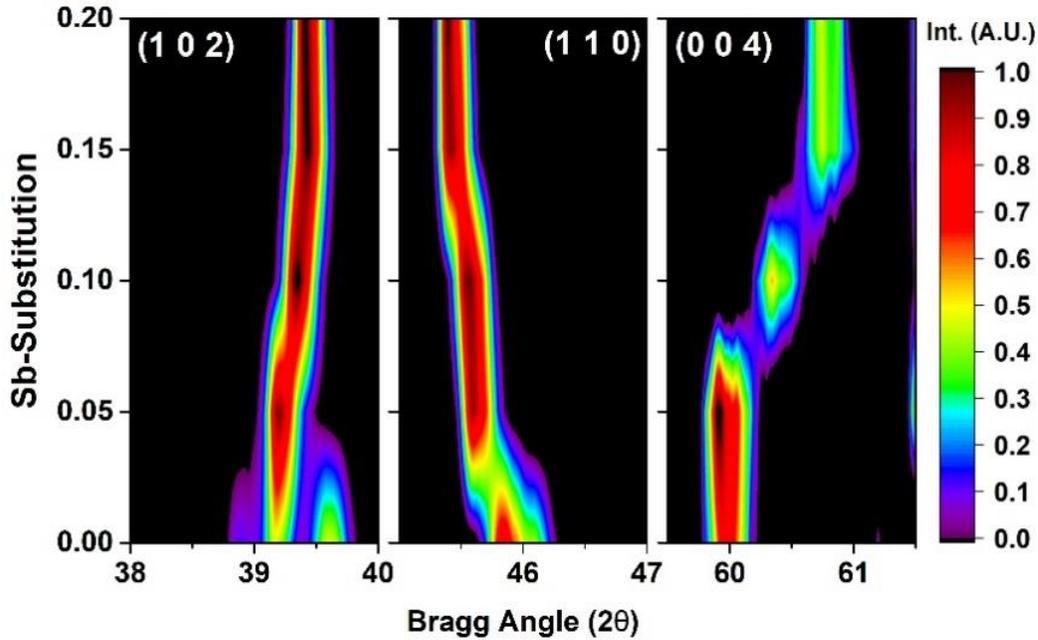

Figure 2: A colorful map of selected diffraction peaks with Miller indices shows the changes in peaks position a long with increasing the Sb substitution ratio in the $CrTe_{1-x}Sb_x$ samples, the intensity of the peaks increases by going from violet to dark red color. The y-axis increases along with the Sb concentration.

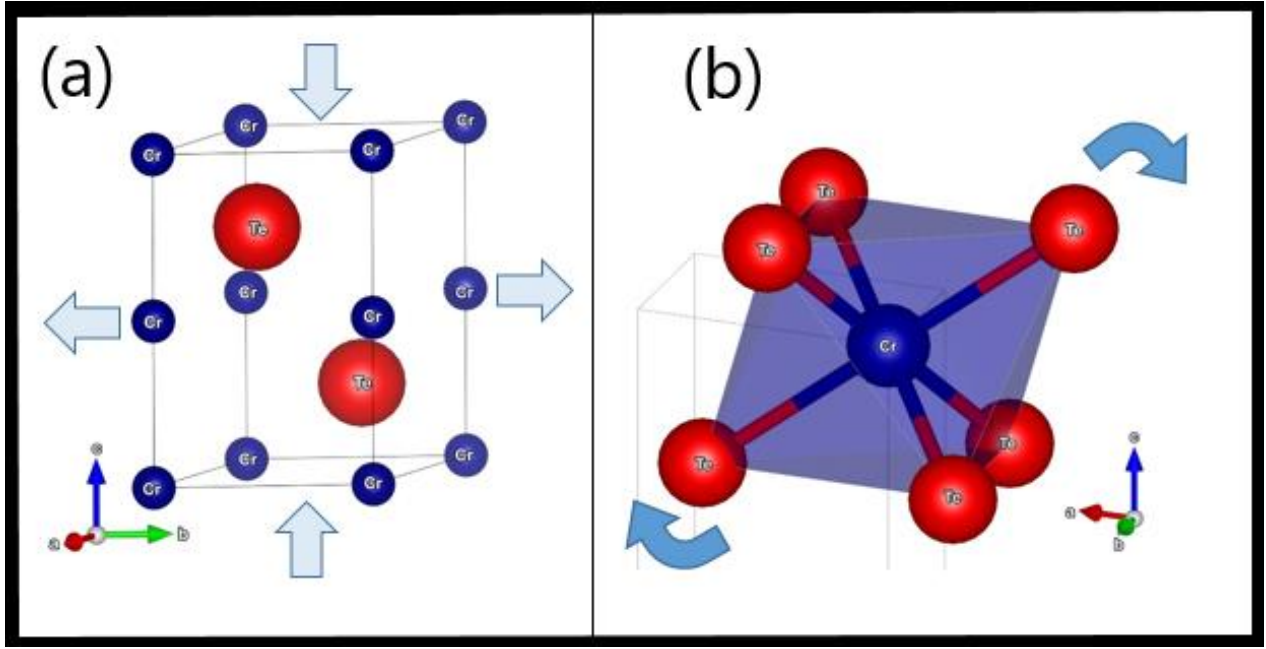

*Figure 3: The unit cell structure of the CrTe1-xSbx hexagonal phase. The red balls (big) represent Te atoms, while the blue ones (small) represent Cr atoms. the arrows at (a) indicates the changes in the unit cell dimensions by increasing the Sb substitution ratio, (b) shows the CrTe octahedron.*

The magnetization dependence on temperature in the range of 50-400 K at H=0.03Tesla is published elsewhere [14]. The thermomagnetic curves for the selected samples (x = 0.00, 0.10, and 0.20) reveals a clear FM-PM phase transition at an inflection points ~ 325.0, 296.0, and 287.0 K respectively [14]. The results also reveal a slight increase in the magnetization along with sharper transition increasing the concentration of antimony. This probably indicates an increase in the effective magnetic moment per *Cr-ion* resulting from the exchange interaction with increasing Sb concentration. Fig. 4 presents the variations of the inverse magnetic susceptibility ($\chi^{-1}$) with temperature. In the paramagnetic region (i.e above Tc), the inverse magnetic susceptibility obey Curie-Weiss law $\chi = C(T - \theta_p)^{-1}$, where C and $\theta_p$ are the Curie constant and the Curie temperature respectively, following a linear behavior down to $\theta_p$ in all samples. The linear extrapolation gives the following ($\theta_p$) values 344, 309, and 302 K for x=0.0, 0.1 and 0.2 respectively [14]. The values of $\theta_p$ are higher than the Tc values obtained from the inflection points of the M vs. T curves. This probably suggests more complex spin configuration than normal FM spin configuration. Recently *Kong et al* observed similar results in CrSbSe$_3$ [37].

The linear portion of the $\chi^{-1}$ vs. T curves can be used to evaluate the effective magnetic moment $\mu_{eff}$ using $\mu_{eff} = \sqrt{3k_B C/N} = \sqrt{8C}\ \mu_B$, where C is the inverse of the slope and $\mu_B$ is the Bohr magneton. The obtained values are $3.92\mu_B$, 4.40 and 4.56 for x=0.0, 0.1 and 0.2 respectively. The expected value for $Cr^{3+}$ is $3.87\mu_B$ [24, 25]. These values indicate a gradual increase in the Cr-effective moment with increasing Sb concentration. These values are in line with the noticed increase in the observed magnetization seen in [14]. The obtained values of $\mu_{eff}$ along with $\theta_p$ are listed in Table 2.

*Table 2: The effective magnetic moments and $\theta_p$ values for the CrTe$_{1-x}$Sb$_x$.*

|  | x=0.00 | x=0.10 | x=0.20 |
|---|---|---|---|
| $\theta_p$(K) | 342.(9) | 308.(3) | 301.(0) |
| $\mu_{eff}$ ($\mu_B$) | 3.9(2) | 4.4(0) | 4.5(6) |

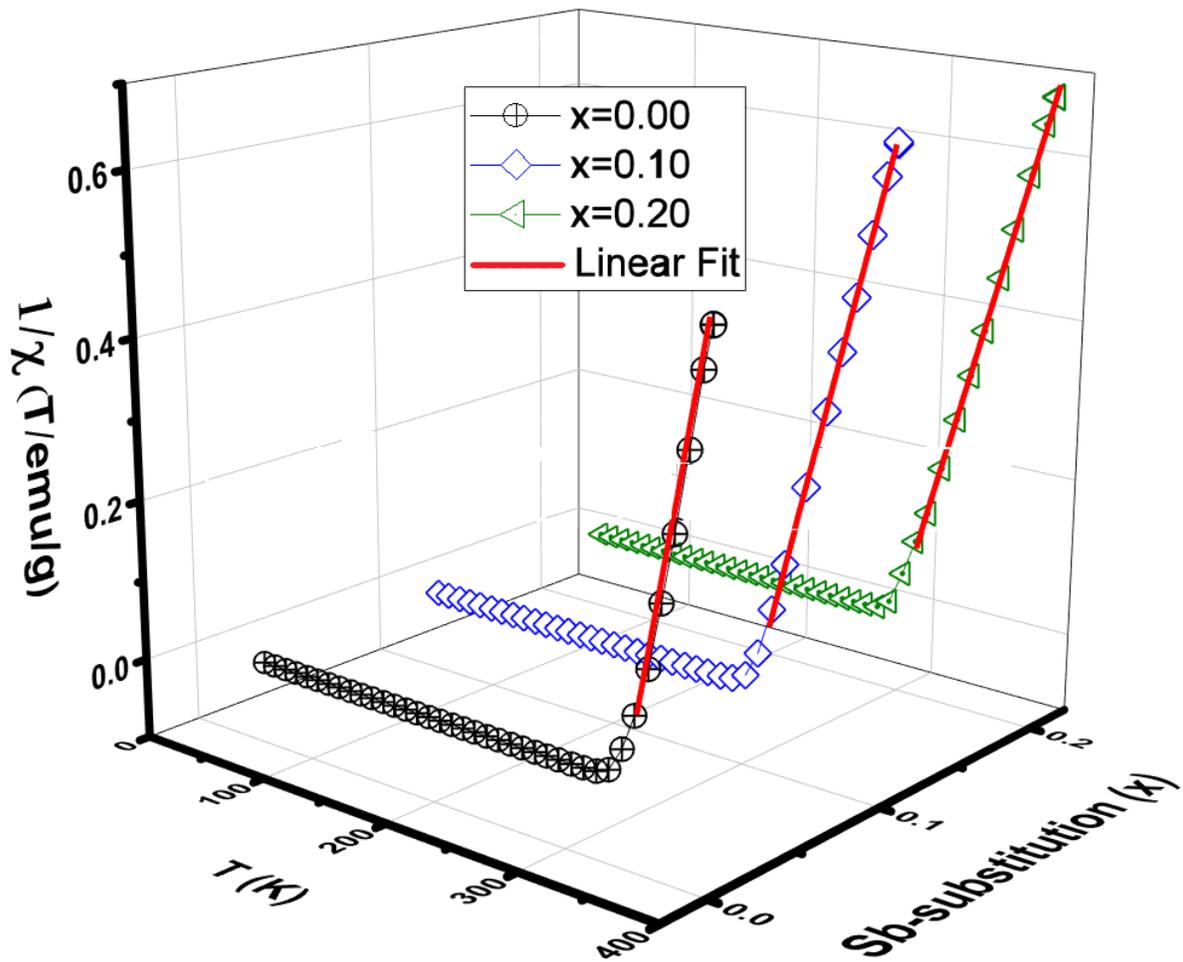

*Figure 4: Temperature-dependent inverse magnetization for different $CrTe_{1-x}Sb_x$ samples measured at H= 0.03 T.*

The magnetic isotherms *M(H, T)* of *CrTe* sample are shown in figure 5 in the temperature range of 300-376 K. The magnetic behavior in Fig. 5 is deeply discussed elsewhere [14]. The magnetization data in Fig. 5 are presented in modified Arrott-plots (MAP) in Fig. 6 in order to investigate the critical behavior and other magnetic properties in the following sections [13, 38-39].

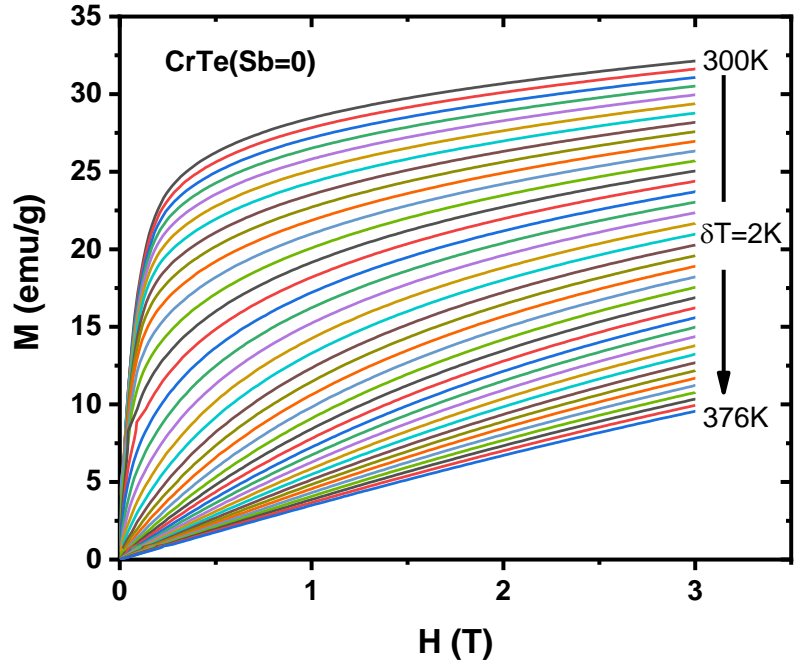

*Figure 5: Variations of the magnetization with field with ΔT=2K intervals for CrTe sample.*

### 4- Critical behaviors

The modified Arrott-plots (MAP) of the magnetization along with Kouvel-Fischer (*K-F*) plots greatly refines the critical exponents near the phase transition [40-41]. The K-F analysis is based on obtaining a linear behavior of the inverse the magnetic susceptibility ($\chi_0^{-1}$) and the spontaneous magnetization (*Ms*). Moreover; *Ms* and $\chi_0^{-1}$ linear graphs follows a universal behavior and their slopes determine the critical exponents [42]. *Ms* and $\chi_0^{-1}$ can be obtained from:

$$M_s(T) = M_0(-\epsilon)^\beta, \quad \text{below } T_c \tag{1}$$

$$\chi_0^{-1} = (h_0/m_0)\epsilon^\gamma, \quad \text{above } T_c \tag{2}$$

$$M = DH^{1/\delta}, \quad \text{at } T_c \tag{3}$$

Where β, γ, and δ are the critical exponents; $\epsilon$ is the reduced temperature; $D$, $M_0$ and $h_0/m_0$ are the critical amplitudes and $M$ is the critical isotherm $M(H)$ at Tc [43]. The critical exponents $\beta, \gamma$ take different values in different magnetic models.

According to the MAP analysis, the magnetic isotherms at high fields are expected to be parallel lines for the correct set of critical exponents. The analysis is based on Arrott-Noakes equation of state; namely:

$$(H/M)^{1/\gamma} = (T - T_C)/A + (M/B)^{1/\beta} \qquad (4)$$

where, $A$ and $B$ are constants [44].

The commonly used magnetic models are: the tricritical mean-field (β = 0.25; γ = 1), 3D-Heisenberg (β = 0.365; γ = 1.336) and the 3D-Ising (β = 0.325; γ = 1.24) [44].

In figure 6 we present the variation of $M^{1/\beta}$ with $(H/M)^{1/\gamma}$ for *CrTe* sample using various magnetic models. The initial slopes of the curves in Fig 6 are positive above and below Tc, according to Banerjee; the positive slopes indicate a second-order phase transition [45]. The slopes at high fields in figure 6 can be used to obtain the relative slope (RS) analysis that allows us to determine the most suitable magnetic model that closely represents the magnetic isotherms. The RS values obtained from Fig 6 are represented in Fig. 7. The RS curve obtained from mean-field model is the closest to the horizontal line at RS = 1 above and below Tc indicating that the mean field model closely represents *CrTe* magnetic state [46]. Similar analysis has been applied for *CrTe$_{1-x}$Sb$_x$* samples. We found that the mean field model is closely represents the magnetic isotherms in the modified Arrott plot representation. The RS curves for *CrTe$_{1-x}$Sb$_x$* samples are presented in Fig. 7 and the modified Arrott plots using the mean field theory (MFT) (for x=0.10 and 0.20) are shown in Fig. 8. It is worth mentioning that, for x=0.20, there are noticeable differences between the previously published data and results obtained in this work (see [30] for comparison). Such differences may be caused by the better quality of the samples used in the current study and the observed relaxation of the atomic-bond in the (*a-b*)*-plane* especially at high Sb-concentration.

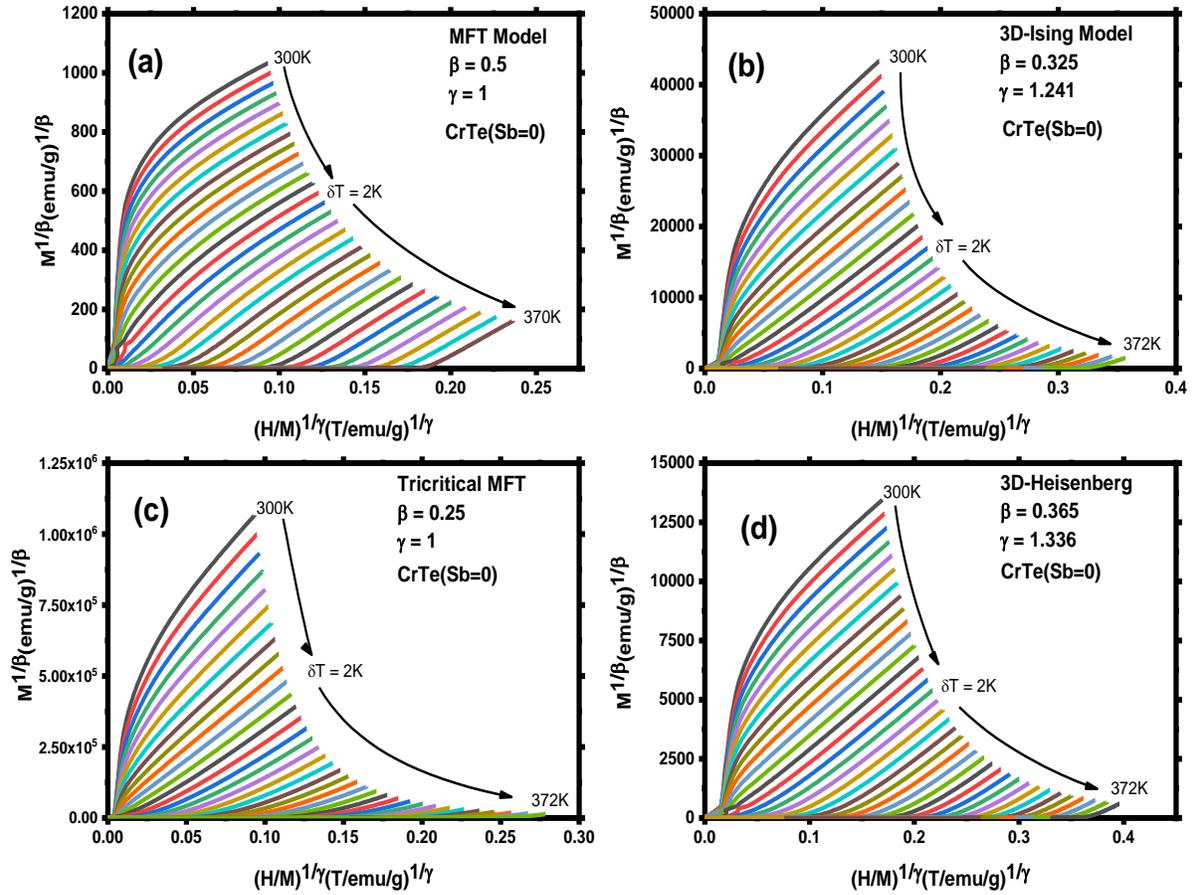

*Figure 6: Modified Arrott-plots for CrTe: (a) Mean-Field. (b) 3D-Ising. (c) Tricritical mean field and (d) 3D-Heisenberg model.*

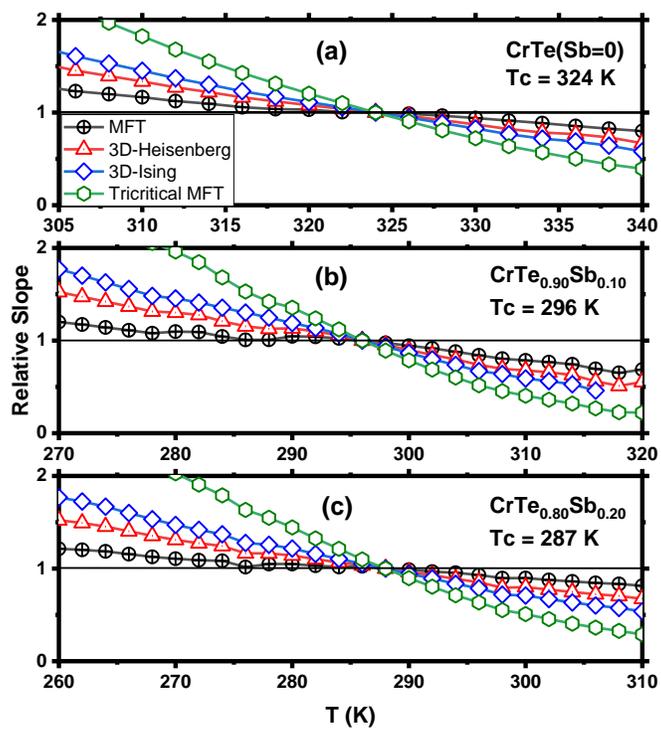

Figure 7: Variations of the relative slope (RS) with temperature for $CrTe_{1-x}Sb_x$ : (a) x=0.0, (b) x=0.1 and (c) x=0.2 samples.

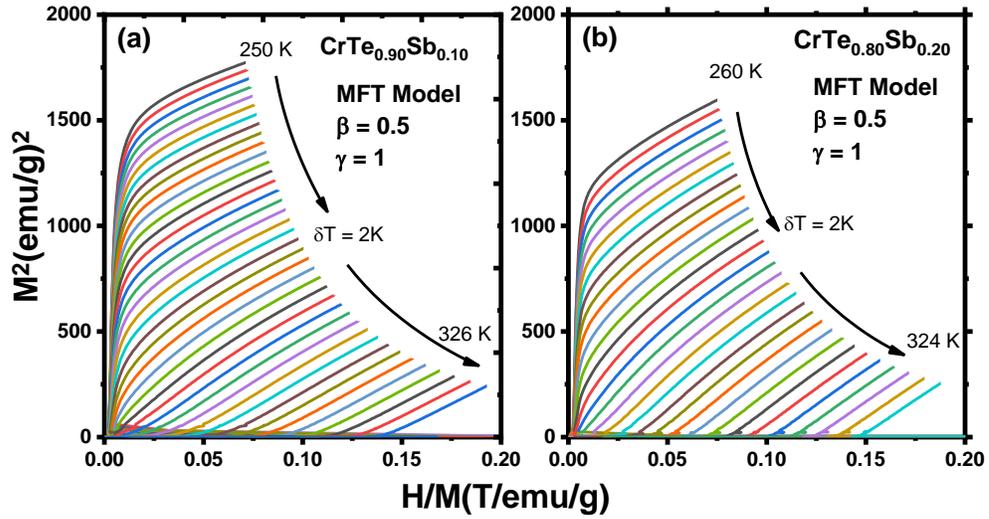

*Figure 8: The Modified Arrott plots using MFT for CrTe$_{1-x}$Sb$_x$ (x=0.1 and 0.2).*

The mean field representation of the modified Arrott plots (Fig. 6 and 8) have been used to extract the spontaneous magnetization $M_s(T)$ and the inverse of the magnetic susceptibility $\chi_0^{-1}(T)$. At each temperature; the curves at high fields are linearly fitted to Eq. 4 and extended to the $M^{1/\beta}$ and $(H/M)^{1/\gamma}$ axes. The intercept points represent the $M_s(T)$ and $\chi_0^{-1}$. The variations of $M_s(T)$ and $\chi_0^{-1}(T)$ with temperature for *CrTe$_{1-x}$Sb$_x$* samples with x=0.0, 0.1 and 0.2 are presented in Figs. 9a, 10a and 11a respectively, and then fitted to Eqs. 1 and 2 in order to evaluate the critical exponents β and γ along with the transition temperature Tc. The obtained values for *CrTe* sample are: $\beta = 0.49 \pm 0.10$, $and\ \gamma = 1.04 \pm 0.02$ and Tc = 334.0±0.1. Similarly, for *CrTe$_{1-x}$Sb$_x$* with x=0.1 and 0.2, the values are: ($\beta = 0.50 \pm 0.01, \gamma = 0.99 \pm 0.01$ and Tc=303.3 ± 0.1) and ($\beta = 0.51 \pm 0.01, \gamma = 1.00 \pm 0.01$ and Tc = 295.4 ± 0.1) respectively.

The Kouvel-Fisher (K-F) plots representations of the spontaneous magnetization and the susceptibility greatly improves the accuracy of modified Arrott analysis for the critical exponents. According to the (K-F) model [41],

$$\frac{M_S(T)}{dM_S(T)/dT} = \frac{T-T_c}{\beta} \tag{5}$$

$$\frac{\chi_0^{-1}(T)}{d\chi_0^{-1}(T)/dT} = \frac{T-T_c}{\gamma} \tag{6}$$

Both equations are linear in temperature ($T-T_c$) with slopes equal to $1/\beta$ and $1/\gamma$. Figs. 9b, 10b and 11b illustrate the K-F-plot for $CrTe_{1-x}Sb_x$ samples with x=0.0, 0.1 and 0.2 respectively. The obtained critical exponents for x=0.0 are: $\beta = 0.49 \pm 0.01 \; and \; \gamma = 0.94 \pm 0.02$ with Tc = 334.4±0.3K. We notice that β is very close to the values obtained using the modified Arrott plot while γ-value has been slightly reduced but still near the theoretical meant-field value. Similar results have been obtained for x=0.10 and x=0.20, these are shown in Fig. 10b and Fig. 11b. All values of the critical exponents extracted from MAP as well as K-F are given in Table 3. Moreover, to obtain more accurate critical exponents (β, and γ) and the critical temperature Tc, we have adopted the iterative method for the unsubstituted sample (CrTe) based on K-F method (Eqns 5 and 6). The results shows, after four rounds, that the values of the critical exponents with Tc are stable and close to the standard mean field model, namely β=0.5, and γ=1, within the uncertainty (see table 4).

The values of third critical exponent (δ) can be calculated using the Widom scaling relation: $\delta = 1 + \frac{\gamma}{\beta}$, and β and γ values obtained from MAP and K-F method [47]. The values are listed in Table 3.

To illustrate the validity of the K-F analysis and the confidence in the obtained critical exponents and critical behavior, we used these critical exponents to re-plot the modified Arrott plots in Figure 12. The figure shows a set of nearly perfect parallel straight lines at high field (H > 2T), indicating that the critical exponents β and γ values obtained using K-F method for the mean field model are quite accurate choice. Moreover, the dashed lines in the Fig. 12 are a linear fitting of isothermal magnetization extrapolated near to the origin point (0,0). The obtained Tc values are 334K, 302K and 294K for x= 0, 0.1 and 0.2 respectively. These values are in agreement with the Tc values obtained from K-F analysis. This further indicates that the critical exponents are close agreement with the MFT critical exponents.

Equation 3 should yield a linear graph of the magnetic isotherms at (T=Tc) in a log-log presentation (Fig. 13) with a slope = 1/δ. The linear fit yields δ = 2.95 ± 0.01. The result is consistent with the one obtained from Widom's relation as shown in Table 3.

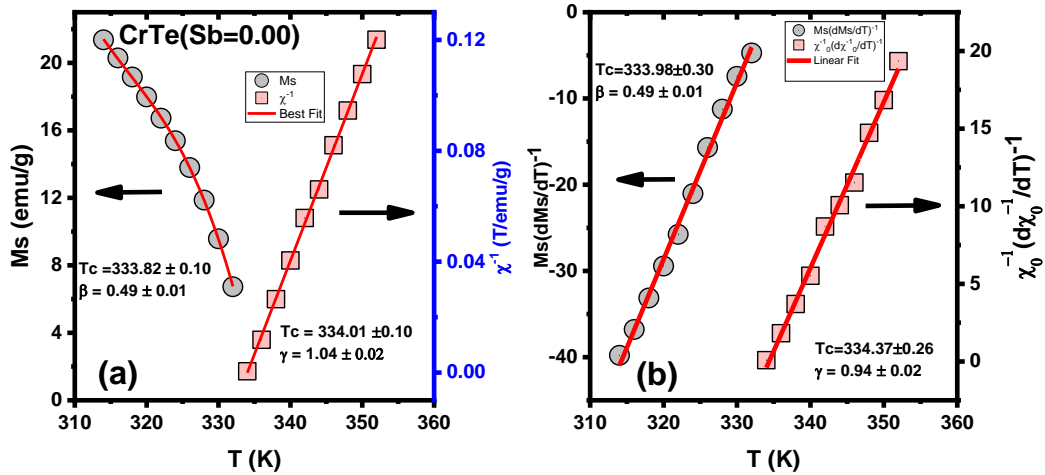

*Figure 9: (a) Variations of $M_s$ and $\chi_0^{-1}$ with temperature, (b) Kouvel-Fisher plots for $M_s(T)$ and $\chi_0^{-1}(T)$ plot for CrTe.*

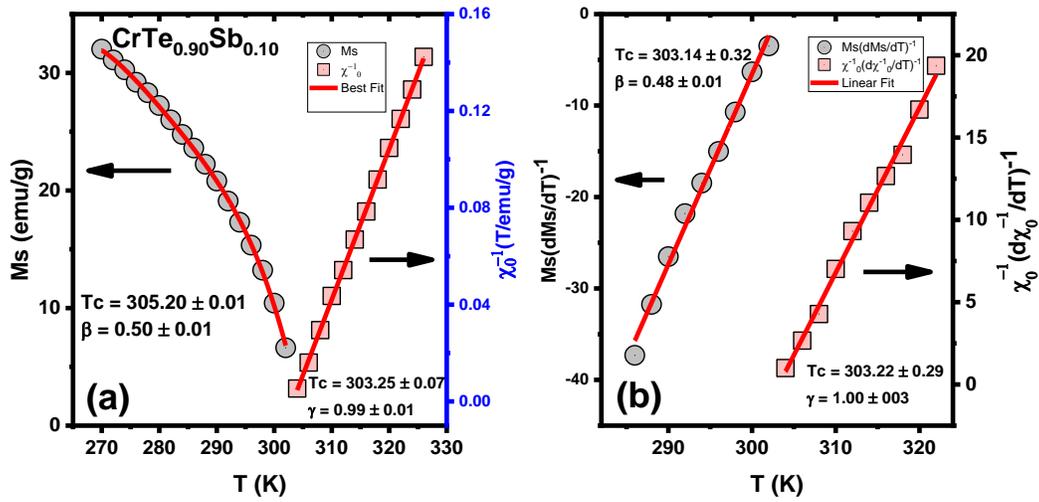

*Figure 10:* *(a) Variations of Ms and $\chi_0^{-1}$ with temperature, (b) Kouvel-Fisher plots for Ms(T) and $\chi_0^{-1}$(T) plot for CrTe$_{0.9}$Sb$_{0.1}$*

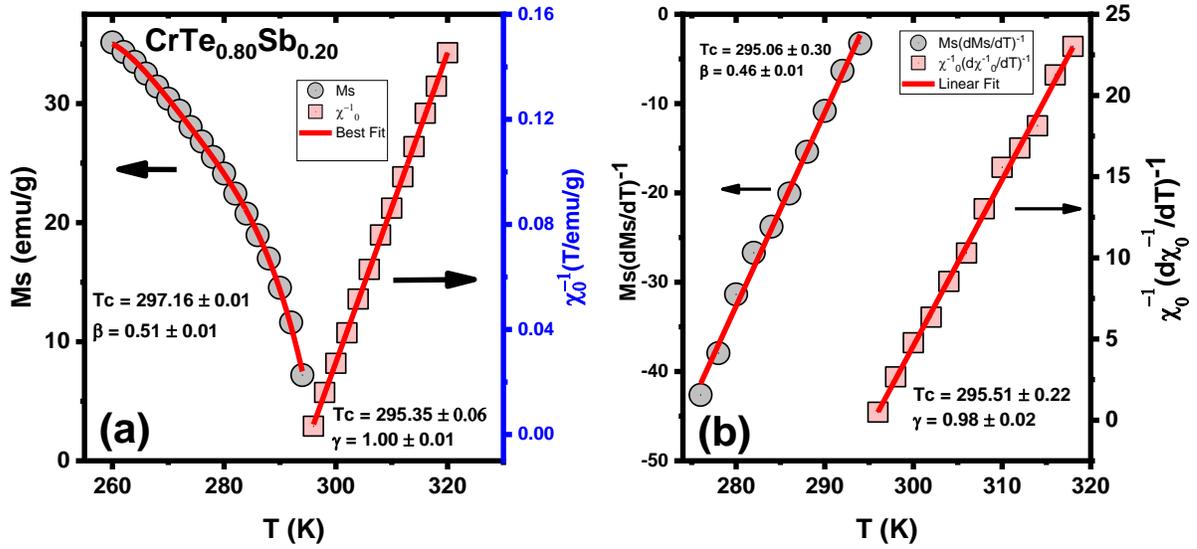

*Figure 11:* *(a) Variations of Ms and $\chi_0^{-1}$ with temperature, (b) Kouvel-Fisher plots for Ms(T) and $\chi_0^{-1}$(T) plot for CrTe$_{0.8}$Sb$_{0.2}$*

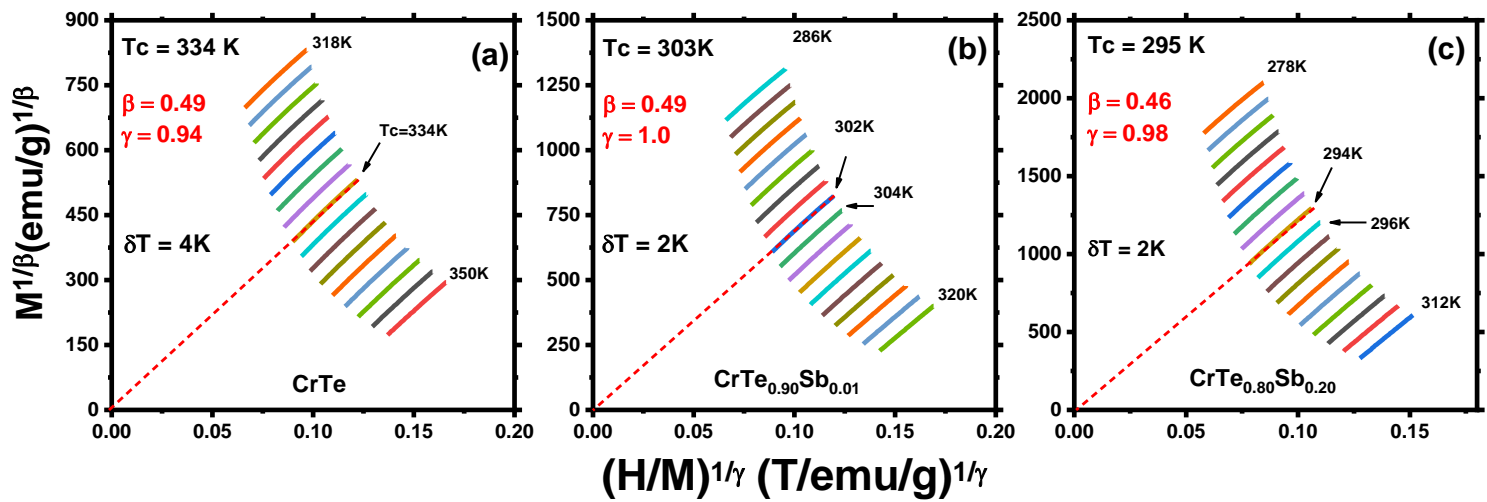

*Figure 12: Modified Arrott plot of $M^{1/\beta}$ vs $(H/M)^{1/\gamma}$ at high field values with the obtained critical exponents of β and γ with a red linear dashed curve passed through the origin for. (a) CrTe, (b) $CrTe_{0.90}Sb_{0.10}$, and (c) $CrTe_{0.80}Sb_{0.20}$*

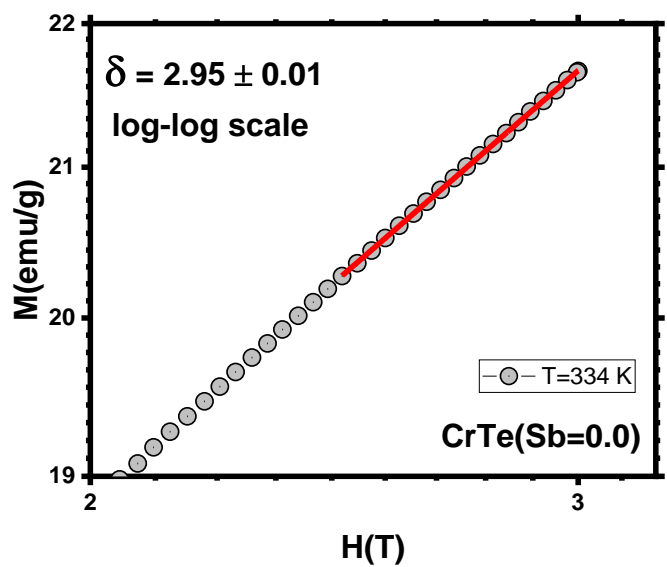

*Figure 13: Critical magnetization isotherm at T=Tc for CrTe in log-log scale.*

*Table 3: Critical exponents along with Tc for CrTe1-xSbx (0.0 ≤ x ≤ 0.2) samples.*

| Sample | Model | Tc(K) | β | γ | δ Theory | δ expt. |
|---|---|---|---|---|---|---|
| CrTe | MAP | 334.0±0.1 | 0.49±0.01 | 1.04±0.02 | 3.12 ± 0.12 | 2.95 ± 0.01 |
| | K-F | 334.4±0.3 | 0.49±0.01 | 0.94±0.02 | 2.92± 0.12 | |
| $CrTe_{0.90}Sb_{0.10}$ | MAP | 303.3±0.1 | 0.50±0.01 | 0.99±0.01 | 2.98±0.09 | 3.11 ± 0.01 |
| | K-F | 303.2±0.3 | 0.48±0.01 | 1.00±0.03 | 3.08 ± 0.16 | |
| $CrTe_{0.80}Sb_{0.20}$ | MAP | 295.4±0.1 | 0.51±0.01 | 1.00±0.01 | 2.96 ± 0.09 | 2.90 ± 0.01 |
| | K-F | 295.5±0.2 | 0.46±0.01 | 0.98±0.02 | 3.13 ± 0.13 | |

*Table 4: Critical exponents along with Tc for CrTe sample using the iterative method*

| Iteration # | Initials | | Tc (K) | β | γ |
|---|---|---|---|---|---|
| | β | γ | | | |
| 0 | 0.5 | 1 | 334.37±0.26 | 0.49±0.01 | 0.94±0.02 |
| 1 | 0.49 | 0.94 | 334.63±0.52 | 0.48±0.01 | 0.96±0.03 |
| 2 | 0.48 | 0.96 | 333.95±0.53 | 0.48±0.01 | 0.97±0.03 |
| 3 | 0.48 | 0.97 | 333.75±0.54 | 0.48±0.01 | 0.98±0.03 |

Close to the transition temperature Tc, the normalized equation of state can be expressed as:

$$m = f_{\pm}(h) \tag{7}$$

Where; $f_{\pm}$ are functions above (+) and below (-) Tc; $m \equiv |\varepsilon|^{-\beta} M(H, \varepsilon)$ is the normalized magnetization and $h \equiv H|\varepsilon|^{-(\beta+\gamma)}$ is the normalized field.

For the correct values of the critical exponents, Eq. (7) implies that for the reliable values of $\beta, \gamma$ and $\delta$; two universal branches above and below Tc presented as $M/|\varepsilon|^\beta$ vs. $H/|\varepsilon|^{(\beta+\gamma)}$ [46]. Both curves merge asymptotically at T=Tc. The critical exponents (β, γ and δ) obtained from Kouvel-Fisher analysis have used to represent the normalized magnetization isotherm (*m*) versus the normalized field (*h*) for all investigated samples. The scaled data for *CrTe* sample is presented in Fig. 14 as two branches of the magnetization above and below Tc. At high fields both branches are converging towards each other at Tc. Fig. 14(b) shows the data in log-log scale, which vividly reveals the merge of the data giving further evidence for the reliability of the critical exponents' values within the mean field model. The scaling analysis has been applied for $CrTe_{1-x}Sb_x$ for x=0.10 and x=0.20. The results are shown in Fig. 15 and Fig. 16 respectively. The well-scaled data for x=0.20 in the current work, comparing to the scaled data published in [30] is further confirm the differences and improvements in the experiment conditions.

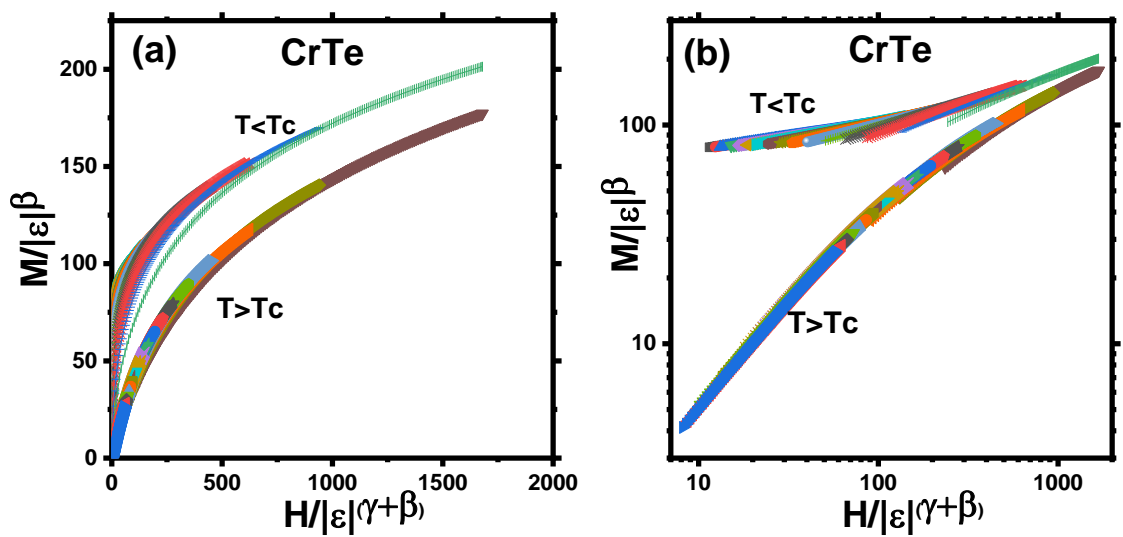

*Figure 14: (a) Universal scaling behavior above and below Tc for CrTe. (b)The log-log scale representation.*

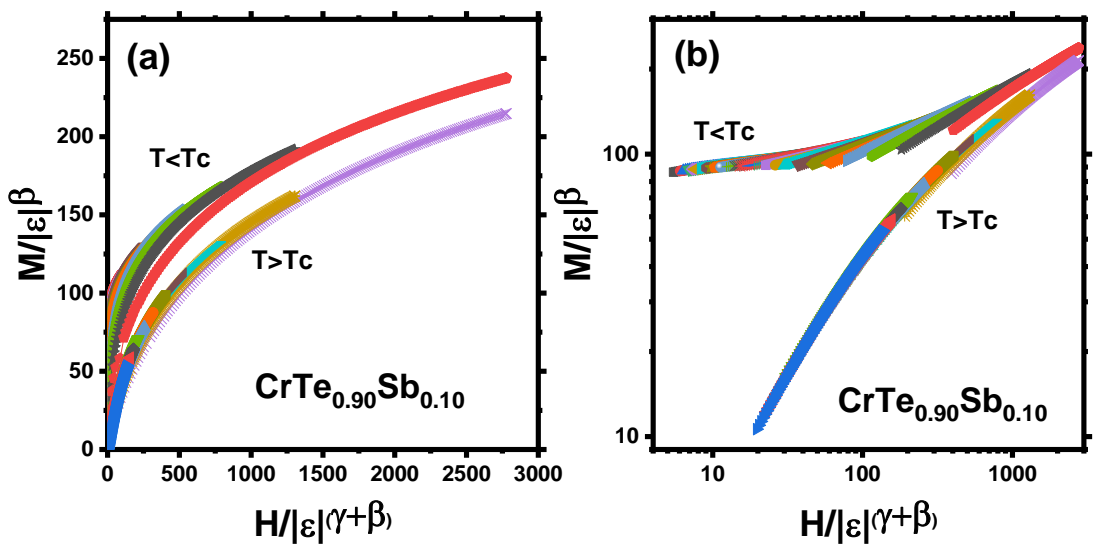

*Figure 15: a) Universal scaling behavior above and below Tc for $CrTe_{0.90}Sb_{0.10}$. (b)The log-log scale representation.*

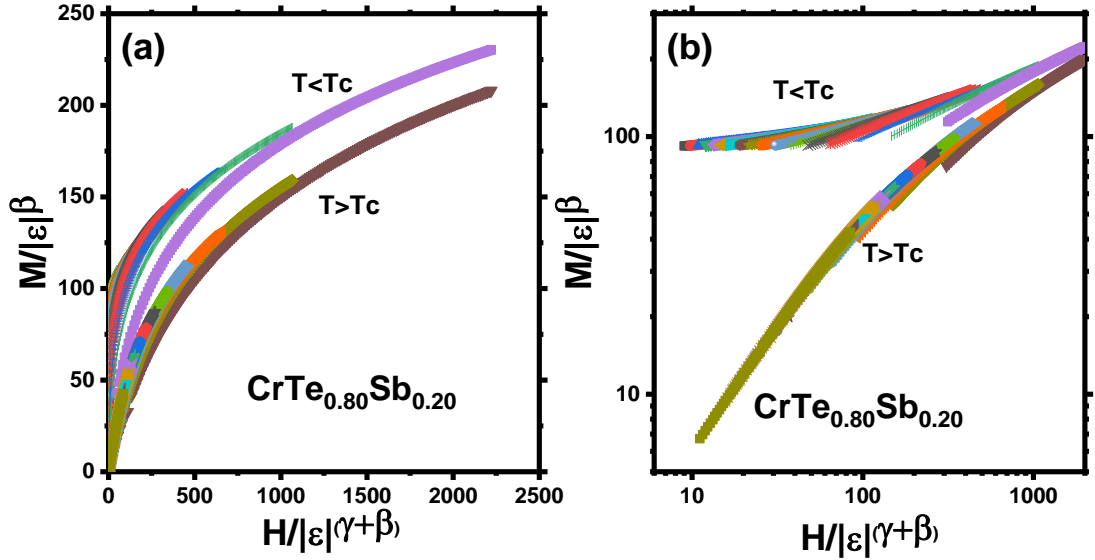

*Figure 16: a) Universal scaling behavior above and below Tc for CrTe$_{0.80}$Sb$_{0.20}$. (b)The log-log scale representation.*

## Conclusion

We investigated the structural and critical behavior of *CrTe$_{1-x}$Sb$_x$* with varying Sb contents. Antimony substitution resulted in pure *NiAs*-hexagonal structure with P6$_3$/mmc (194) space-group. The critical exponents ($\beta$, $\gamma$, and $\delta$) follow mean field model for all samples with different Sb-concentrations. The magnetization isotherms near the ferromagnetic transition follow a universal scaling behavior, giving further support for the mean field behavior in *CrTe$_{1-x}$Sb$_x$* and confirms the obtained values of the critical exponents.

## Acknowledgments

Authors from KFUPM acknowledge the support provided by the Deanship of Scientific Research at King Fahd University of Petroleum & Minerals (KFUPM) for funding this work under project **No. SR191008**.

AAE acknowledges the start-up and rising stars funds by UTEP and UT-system respectively and the partial supported by the US-National Science Foundation under award **No. 2009358**.